\documentclass[amsmath, aps]{revtex4}
\usepackage{graphicx}
\usepackage{times}
\usepackage{epsfig}
\usepackage{amssymb}

\usepackage{amsmath}
\begin{document}
\title{QED calculations of three-photon transition probabilities in H-like ions with arbitrary nuclear charge}
\author{T. Zalialiutdinov$^{1}$, D. Solovyev$^{1}$ and L. Labzowsky$^{1,2}$}

\affiliation{$^1$ Department of Physics, St.Petersburg State University,
Ulianovskaya 1, Petrodvorets, St.Petersburg 198504, Russia
\\
$^2$  Petersburg Nuclear Physics Institute, 188300, Gatchina, St.
Petersburg, Russia}

\begin{abstract}
The quantum electrodynamical theory of the three-photon transitions in hydrogen-like ions is presented. Emission probabilities of various three-photon decay channels for $ 2p_{3/2} $, $ 2p_{1/2} $ and $ 2s_{1/2} $ states are calculated for the nuclear charge $Z$ values $1\leqslant Z \leqslant 95$. The results are given in two different gauges. The fully relativistic three-photon decay rates of hydrogen-like ions with half-integer nuclear spin are given for transitions between fine structure components. The results can be applied to the tests of the Bose-Einstein statistics for the multiphoton systems.
\end{abstract}

\maketitle

\section{Introduction}
During last years multiphoton processes have attracted special attention in different fields of physical science. Multiphoton transitions in atomic systems, e.g., two-photon decay and absorption have become a useful and even a standard tool for experimental studies of diverse spectroscopic characteristics in atoms: excitation of levels of various systems, the determination of physical constants, parity-violation phenomena and etc. Since the primary works of Kramers and Heisenberg \cite{kramers}, Waller \cite{waller}, and Goeppert-Mayer \cite{goeppert}, there has been a continuing interest in accurate calculations of multiphoton transitions in hydrogenlike ions. Two- and three-photon transitions are also of interest in astrophysics \cite{seager}, \cite{chluba}. Moreover, multiphoton transitions has been studied in connection with the test of Bose-Einstein statistics \cite{demille}-\cite{angom}. Recently the Spin-Statistics Selection Rules (SSSRs) which present an extension of the Landau-Yang Theorem (LYT) to the multiphoton processes in atoms were established \cite{zlsg}. Examples given in \cite{zlsg} concerned to the He-like Highly Charged Ions (HCI) with photon frequencies in the X-ray region. In the present work we suggest the experiments for the tests of the 3-photon SSSRs on H-like HCI with the transitions in optical region. 

In present work the QED theory is applied to the study of spontaneous three-photon decay processes. Calculations of the various transition probabilities for H-like ions for nuclear charge $ Z $ values within the region $1 \leqslant Z \leqslant 95$ are performed. Furthermore fully relativistic calculations for three-photon transitions between fine structure components with account for hyperfine structure of hydrogen-like ions are presented. This means that we have fixed a certain hyperfine substate of the fine structure component, not introducing explicitly the hyperfine level splitting. We performed calculations in two different gauges, what allows for the accurate check of the gauge invariance of the results. For the summation over the complete Dirac spectrum the B-spline method \cite{bspline} was employed. Relativistic units are used throughout this paper.

Our paper is organized as follows. In section II we present detailed QED derivation of the general expression for spontaneous three-photon decay rate in H-like ion for an arbitrary combination of electric and magnetic multipoles and in an arbitrary gauge for the electromagnetic potentials. In section III we consider three-photon transitions between fine structure components with account for hyperfine structure (as explained above) of an one-electron ion with arbitrary nuclear spin $I$. Numerical values for the transition probabilities considered in sections II and III are presented in Tables I-III. In section IV we discuss the application of our results to the study of Bose-Einstein statistics and to the extension of LYT \cite{zlsg}-\cite{yang}. This section contains also the concluding remarks. 

\section{QED theory of three-photon transitions}
We present a computationally convenient fully relativistic form of a general expression for the 3-photon decay rate in H-like ion for an arbitrary combination of electric and magnetic multipoles and in an arbitrary gauge for the electromagnetic potentials. Relativistic units $\hbar =c=1$ are employed. In this section the hyperfine structure of the levels is neglected.

The $S$-matrix element for the emission process $i\rightarrow f+3\gamma$ ($i$ and $f$ denote the initial and final states respectively) reads \cite{berlif}, \cite{akhiezer}, \cite{ALPS}
\begin{eqnarray}
\label{11}
S_{fi}^{(3)}=(-\rm{i}e)^3\int d^4x_3d^4x_2d^4x_1\overline{\psi}_{f}(x_3)\gamma_{\mu_3}A_{\mu_3}^{*\left(\vec{k}_{3}\vec{e}_{3}\right)}(x_3)S(x_3,x_2)\gamma_{\mu_2}A_{\mu_2}^{*\left(\vec{k}_{2}\vec{e}_{2}\right)}(x_2) S(x_2,x_1)\gamma_{\mu_1}A_{\mu_1}^{*\left(\vec{k}_{1}\vec{e}_{1}\right)}(x_1)\psi_{i}(x_1)\;,
\end{eqnarray}
where
\begin{eqnarray}
\label{12}
\psi_n(x)=\psi_n(\vec{r})e^{-\rm{i}E_nt},
\end{eqnarray}
$\psi_n(\vec{r})$ is the solution of the Dirac equation for the atomic electron, $E_n$ is the Dirac energy, $\overline{\psi}_n=\psi^+_n\gamma_0$ is the Dirac conjugated wave function, $\gamma_{\mu}\equiv(\gamma_0,\vec{\gamma})$ are the Dirac matrices and $x\equiv(\vec{r},\rm{i}\,t)$ are the space-time coordinates. In this paper the Euclidean metric with an imaginary fourth vector component is adopted. The photon wave function (electromagnetic field potential) is described by
\begin{eqnarray}
\label{13}
A_{\mu}^{(\vec{k},\vec{e})}(x)=\sqrt{\frac{2\pi}{\omega}}e_{\mu}e^{\rm{i}k_{\mu}x_{\mu}}=A_{\mu}^{(\vec{k},\vec{e})}(\vec{r})e^{-\rm{i}\omega t},
\end{eqnarray}
where $k\equiv(\vec{k},\rm{i}\omega)$ is the photon momentum 4-vector, $\vec{k}$ is the photon wave vector, $\omega = |\vec{k}|$ is the photon frequency, $e_{\mu}$ are the components of the photon polarization 4-vector, $\vec{e}$ is the 3-dimensional polarization vector for real photons, $A_{\mu}^{(\vec{k},\vec{e})}$ corresponds to the absorbed photon and $A_{\mu}^{*(\vec{k},\vec{e})}$ represents the emitted photon, respectively. 

For the real transverse photons
\begin{eqnarray}
\label{14}
\vec{A}^{(\vec{k},\vec{e})}(x)=\sqrt{\frac{2\pi}{\omega}}\vec{e}e^{\rm{i}(\vec{k}\vec{r}-\omega t)}\equiv\sqrt{\frac{2\pi}{\omega}}\vec{A}_{\vec{e},\vec{k}} e^{-\rm{i}\omega t}.
\end{eqnarray}

The electron propagator for bound electrons can be presented in the form of the eigenmode decomposition with respect to one-electron eigenstates \cite{berlif}, \cite{akhiezer}
\begin{eqnarray}
\label{15}
S(x_1,x_2)=\frac{1}{2\pi \rm{i}}\int\limits^{\infty}_{-\infty}d\omega e^{\rm{i}\omega(t_1-t_2)}\sum\limits_{n}\frac{\psi_n(\vec{r}_1)\overline{\psi}_n(\vec{r}_2)}{E_n(1-\rm{i}0)+\omega}\;.
\end{eqnarray}
Here summation runs over entire Dirac spectrum for atomic electron. Insertion of the expressions (\ref{12})-(\ref{15}) into Eq. (\ref{11}) and performing the integrations over time and frequency variables yields
\begin{eqnarray}
\label{16}
S_{fi}^{(3)}=-2\pi \rm{i} e^3\delta(E_i-E_f-\omega_3-\omega_2-\omega_1)\sum\limits_{n'n}\frac{\left(\vec{\alpha}\vec{A}^*_{\vec{e}_3,\vec{k}_3}\right)_{fn'}\left(\vec{\alpha}\vec{A}^*_{\vec{e}_2,\vec{k}_2}\right)_{n'n}\left(\vec{\alpha}\vec{A}^*_{\vec{e}_1,\vec{k}_1}\right)_{ni}}{(E_{n'} - E_f - \omega_3) (E_n - E_f - \omega_3 - \omega_2)},
\end{eqnarray}
where $\vec{\alpha}$ are the Dirac matrices, $(\dots)_{kn}$ denotes the matrix element with Dirac wave function $\psi_k$, $\psi_n$. The amplitude of the emission process $ U_{if} $ is related to the $S$-matrix element via
\begin{eqnarray}
\label{17}
S_{if}=-2\pi \rm{i} \delta(E_i-E_f-\omega_3-\omega_2-\omega_1)U_{if}.
\end{eqnarray}
The differential probability (transition rate) of the process is defined as
\begin{eqnarray}
\label{18}
\frac{dW^{3\gamma}_{i\rightarrow f}}{d\omega_3d\omega_2d\omega_1}=2\pi\delta(E_i-E_f-\omega_3-\omega_2-\omega_1)\left|U^{(3)}_{if}\right|^2\;.
\end{eqnarray}
Then the differential transition rate in conjunction with the integration over photon directions $ \vec{\nu}=\vec{k}/|\vec{k}| $ and summation over the photon polarizations $ \vec{e} $ of all the emitted photons with all permutations of photons is
\begin{eqnarray}
\label{19}
\frac{dW_{i\rightarrow f}(\omega_1,\omega_2)}{d\omega_1d\omega_2} = \frac{\omega_3\;\omega_2\;\omega_1}{(2\pi)^5}\sum_{\vec{e}_3,\vec{e}_2,\vec{e}_1}\int d\vec{\nu}_1d\vec{\nu}_2d\vec{\nu}_3 \times
\\
\nonumber
\left| \sum\limits_{n'n}\frac{\left(\vec{\alpha}\vec{A}^*_{\vec{e}_3,\vec{k}_3}\right)_{fn'}\left(\vec{\alpha}\vec{A}^*_{\vec{e}_2,\vec{k}_2}\right)_{n'n}\left(\vec{\alpha}\vec{A}^*_{\vec{e}_1,\vec{k}_1}\right)_{ni}}{(E_{n'} - E_f - \omega_3) (E_n - E_f - \omega_3 - \omega_2)}\right.+
%\\
%\nonumber
\sum\limits_{n'n}\frac{\left(\vec{\alpha}\vec{A}^*_{\vec{e}_3,\vec{k}_3}\right)_{fn'}\left(\vec{\alpha}\vec{A}^*_{\vec{e}_1,\vec{k}_1}\right)_{n'n}\left(\vec{\alpha}\vec{A}^*_{\vec{e}_2,\vec{k}_2}\right)_{ni}}{(E_{n'} - E_f - \omega_3) (E_n - E_f - \omega_3 - \omega_1)}+
\\
\nonumber
\sum\limits_{n'n}\frac{\left(\vec{\alpha}\vec{A}^*_{\vec{e}_1,\vec{k}_1}\right)_{fn'}\left(\vec{\alpha}\vec{A}^*_{\vec{e}_3,\vec{k}_3}\right)_{n'n}\left(\vec{\alpha}\vec{A}^*_{\vec{e}_2,\vec{k}_2}\right)_{ni}}{(E_{n'} - E_f - \omega_1) (E_n - E_f - \omega_1 - \omega_3)}+
%\\
%\nonumber
\sum\limits_{n'n}\frac{\left(\vec{\alpha}\vec{A}^*_{\vec{e}_1,\vec{k}_1}\right)_{fn'}\left(\vec{\alpha}\vec{A}^*_{\vec{e}_2,\vec{k}_2}\right)_{n'n}\left(\vec{\alpha}\vec{A}^*_{\vec{e}_3,\vec{k}_3}\right)_{ni}}{(E_{n'} - E_f - \omega_1) (E_n - E_f - \omega_1 - \omega_2)}+
\\
\nonumber
\sum\limits_{n'n}\frac{\left(\vec{\alpha}\vec{A}^*_{\vec{e}_2,\vec{k}_2}\right)_{fn'}\left(\vec{\alpha}\vec{A}^*_{\vec{e}_3,\vec{k}_3}\right)_{n'n}\left(\vec{\alpha}\vec{A}^*_{\vec{e}_1,\vec{k}_1}\right)_{ni}}{(E_{n'} - E_f - \omega_2) (E_n - E_f - \omega_2 - \omega_3)}+
%\\
%\nonumber
\left.\sum\limits_{n'n}\frac{\left(\vec{\alpha}\vec{A}^*_{\vec{e}_2,\vec{k}_2}\right)_{fm}\left(\vec{\alpha}\vec{A}^*_{\vec{e}_1,\vec{k}_1}\right)_{n'n}\left(\vec{\alpha}\vec{A}^*_{\vec{e}_3,\vec{k}_3}\right)_{ni}}{(E_{n'} - E_f - \omega_2) (E_n - E_f - \omega_2 - \omega_1)}\right|^2,
\end{eqnarray}
where the frequency $\omega_3$ is defines via $\delta$ function in Eq. (\ref{18}). Therefore the total transition rate is
\begin{eqnarray}
\label{21}
W_{i\rightarrow f}=\frac{1}{3!}\frac{1}{2j_{ei}+1}\sum\limits_{m_{ei},m_{ef}}\;\iint \frac{dW_{i\rightarrow f}(\omega_1,\omega_2)}{d\omega_1d\omega_2}d\omega_1d\omega_2\;,
\end{eqnarray}
where $ j_{ei} $, $ m_{ei} $, $ j_{ef} $, $ m_{ef} $ are the angular momenta and their projections for the initial $ (i) $ and final $ (f) $ electron states.

Expanding the plane waves into spherical waves in Eq. (\ref{19}) we go over to the description of photons by the total angular momentum  $j$, its projection $m$ and parity (type of the photon). Then we arrive at 
\begin{eqnarray}
\label{20}
\frac{dW_{i\rightarrow f}(\omega_1,\omega_2)}{d\omega_1d\omega_2} = \frac{\omega_3\;\omega_2\;\omega_1}{(2\pi)^5}\sum_{\lambda_3\lambda_2\lambda_1}\sum_{j_{\gamma_3}j_{\gamma_2}j_{\gamma_1}}\sum_{m_{\gamma_3}m_{\gamma_2}m_{\gamma_1}}
\\
\nonumber
\left|\sum\limits_{n'n}\frac{\left(Q^{(\lambda_3)}_{j_{\gamma_3}m_{\gamma_3}\omega_3}\right)_{fn'}\left(Q^{(\lambda_2)}_{j_{\gamma_2}m_{\gamma_2}\omega_2}\right)_{n'n}\left(Q^{(\lambda_1)}_{j_{\gamma_1}m_{\gamma_1}\omega_1}\right)_{ni}}{(E_{n'} - E_f - \omega_3) (E_n - E_f - \omega_3 - \omega_2)}+
%\right.+
%\\
%\nonumber
\sum\limits_{n'n}\frac{\left(Q^{(\lambda_3)}_{j_{\gamma_3}m_{\gamma_3}\omega_3}\right)_{fn'}\left(Q^{(\lambda_1)}_{j_{\gamma_1}m_{\gamma_1}\omega_1}\right)_{n'n}\left(Q^{(\lambda_2)}_{j_{\gamma_2}m_{\gamma_2}\omega_2}\right)_{ni}}{(E_{n'} - E_f - \omega_3) (E_n - E_f - \omega_3 - \omega_1)}+
\right.
\\
\nonumber
\sum\limits_{n'n}\frac{\left(Q^{(\lambda_1)}_{j_{\gamma_1}m_{\gamma_1}\omega_1}\right)_{fn'}\left(Q^{(\lambda_3)}_{j_{\gamma_3}m_{\gamma_3}\omega_3}\right)_{n'n}\left(Q^{(\lambda_2)}_{j_{\gamma_2}m_{\gamma_2}\omega_2}\right)_{ni}}{(E_{n'} - E_f - \omega_1) (E_n - E_f - \omega_1 - \omega_3)}+
%\\
%\nonumber
\sum\limits_{n'n}\frac{\left(Q^{(\lambda_1)}_{j_{\gamma_1}m_{\gamma_1}\omega_1}\right)_{fn'}\left(Q^{(\lambda_2)}_{j_{\gamma_2}m_{\gamma_2}\omega_2}\right)_{n'n}\left(Q^{(\lambda_3)}_{j_{\gamma_3}m_{\gamma_3}\omega_3}\right)_{ni}}{(E_{n'} - E_f - \omega_1) (E_n - E_f - \omega_1 - \omega_2)}+
\\
\nonumber
\sum\limits_{n'n}\frac{\left(Q^{(\lambda_2)}_{j_{\gamma_2}m_{\gamma_2}\omega_2}\right)_{fn'}\left(Q^{(\lambda_3)}_{j_{\gamma_3}m_{\gamma_3}\omega_3}\right)_{n'n}\left(Q^{(\lambda_1)}_{j_{\gamma_1}m_{\gamma_1}\omega_1}\right)_{ni}}{(E_{n'} - E_f - \omega_2) (E_n - E_f - \omega_2 - \omega_3)}+
%\\
%\nonumber
\left.\sum\limits_{n'n}\frac{\left(Q^{(\lambda_2)}_{j_{\gamma_2}m_{\gamma_2}\omega_2}\right)_{fn'}\left(Q^{(\lambda_1)}_{j_{\gamma_1}m_{\gamma_1}\omega_1}\right)_{n'n}\left(Q^{(\lambda_3)}_{j_{\gamma_3}m_{\gamma_3}\omega_3}\right)_{ni}}{(E_{n'} - E_f - \omega_2) (E_n - E_f - \omega_2 - \omega_1)}\right|^2\;.
\end{eqnarray} 
In Eq. (\ref{20}) we employ the reduction of the matrix elements $ \left(Q\right)_{ab} $ to the radial integrals developed in \cite{grant}, \cite{goldman}
\begin{eqnarray}
\label{22}
\left(Q^{(\lambda)}_{j_{\gamma}m_{\gamma}\omega}\right)_{n_{\alpha}j_{e\alpha}l_{e\alpha}m_{e\alpha},n_{\beta}j_{e\beta}l_{e\beta}m_{e\beta}}=(-1)^{j_{e\alpha}-m_{e\alpha}}
\begin{pmatrix}
j_{e\alpha} & j_{\gamma} & j_{e\beta}\\
-m_{e\alpha} & m_{\gamma} & m_{e\beta}
\end{pmatrix}\times
\\
\nonumber
(-\rm{i})^{j_{\gamma}+\lambda-1}(-1)^{j_{e\alpha}-1/2}\left(\dfrac{4\pi}{2j_{\gamma}+1}\right)^{1/2}
\left[(2j_{e\alpha}+1)(2j_{e\beta}+1)\right]^{1/2}
\begin{pmatrix}
j_{e\alpha} & j_{\gamma} & j_{e\beta}\\
1/2 & 0 & -1/2
\end{pmatrix}
\overline{M}^{(\lambda,j_{\gamma})}_{n_{\alpha}l_{e\alpha}n_{\beta}l_{e\beta}}(\omega).
\end{eqnarray}
Here $j_{\gamma}$, $m_{\gamma}$ are the total angular momentum of the photon and its projection, $\lambda$ characterizes the type of the photon: $\lambda=1$ corresponds to electric and $\lambda=0$ corresponds the magnetic photons. The indices $n_{\alpha},j_{e\alpha},l_{e\alpha},m_{e\alpha}$ present a standard set of one-electron Dirac quantum numbers. The radial matrix elements $\overline{M}^{(\lambda,j_{\gamma})}_{n_{\alpha}l_{e\alpha}n_{\beta}l_{e\beta}}$ in Eq. (\ref{22}) are equal to
\begin{eqnarray}
\label{23}
\overline{M}^{(1,j_{\gamma})}_{n_{\alpha}l_{\alpha}n_{\beta}l_{\beta}}=\left[\left(\frac{j_{\gamma}}{j_{\gamma}+1}\right)^{1/2}\left[\left(\kappa_{\alpha}-\kappa_{\beta}\right)I^{+}_{j_{\gamma}+1}+(j_{\gamma}+1)I^{-}_{j_{\gamma}+1}\right]-\left(\frac{j_{\gamma}+1}{j_{\gamma}}\right)^{1/2}\left[\left(\kappa_{\alpha}-\kappa_{\beta}\right)I^{+}_{j_{\gamma}-1}-j_{\gamma}I^{-}_{j_{\gamma}-1}\right]\right]
\\
\nonumber
 - G\left[(2j_{\gamma}+1)J_{j_{\gamma}}+\left(\kappa_{\alpha}-\kappa_{\beta}\right)\left(I^{+}_{j_{\gamma}+1}-I^{+}_{j_{\gamma}-1}\right)-j_{\gamma}I^{-}_{j_{\gamma}-1}+\left(j_{\gamma}+1\right)I^{-}_{j_{\gamma}+1}\right],
\end{eqnarray}
\begin{eqnarray}
\label{24}
\overline{M}^{(0,j_{\gamma})}_{n_{\alpha}l_{\alpha}n_{\beta}l_{\beta}}=\frac{2j_{\gamma}+1}{\left[j_{\gamma}(j_{\gamma}+1)\right]^{1/2}}\left(\kappa_{\alpha}+\kappa_{\beta}\right)I^{+}_{j_{\gamma}},
\end{eqnarray}
\begin{eqnarray}
\label{25}
I^{\pm}_{j_{\gamma}}=\int\limits_{0}^{\infty}\left(g_{\alpha}f_{\beta}\pm f_{\alpha}g_{\beta}\right)j_{j_{\gamma}}\left(\omega r\right)dr,
\end{eqnarray}
\begin{eqnarray}
\label{26}
J_{j_{\gamma}}=\int\limits_{0}^{\infty}\left(g_{\alpha}g_{\beta}+ f_{\alpha}f_{\beta}\right)j_{j_{\gamma}}\left(\omega r\right)dr\;,
\end{eqnarray}
where $g_{\alpha}$ and $f_{\alpha}$ are the large and small components of the radial Dirac wave function as defined in \cite{grant}, $\kappa$ is the Dirac angular number, $\omega$ is the photon frequency, $ j_{j_{\gamma}} $ represents the spherical Bessel function, $G$  is the gauge parameter for the electromagnetic potentials. In our calculations we employ the "velocity" gauge $\left(G=0\right)$ and the "length" $\left(G=\sqrt{\frac{j_{\gamma}+1}{j_{\gamma}}}\right)$ gauge for the matrix element Eq. (\ref{23}) \cite{labzsol}. Note that equation (\ref{14}) corresponds to $\left(G=0\right)$.

The results can be further simplified by the summations over projections of all the angular momenta. For this purpose we define the radial integral part for a particular combination of multipoles as
\begin{eqnarray}
\label{27}
S^{j_{n'}j_{n}}(i,j,k)=\sum_{l_{en'},l_{en}}\sum_{n,n'}\frac{\overline{M}^{(\lambda_i,j_{\gamma_i})}_{f,n'}(\omega_i)\overline{M}^{(\lambda_j,j_{\gamma_j})}_{n',n}(\omega_j)\overline{M}^{(\lambda_k,j_{\gamma_k})}_{n,i}(\omega_k)}{\left(E_{n'j_{en'}l_{en'}}-E_{n_fj_{ef}l_{ef}}-\omega_k-\omega_j\right)\left(E_{nj_{en}l_{en}}-E_{n_fj_{ef}l_{ef}}-\omega_k\right)}
\times
\\
\nonumber
\Delta^{j_{en'},j_{en}}\;\pi^{l_{en'}}_f(i)\;\pi^{l_{en}}_{n'}(j)\;\pi^{l_{ei}}_{n}(k),
\end{eqnarray}
where 
\begin{eqnarray}
\pi^{l}_k(t)=
\left\{
	\begin{array}{l l}
		1\;\mbox{if}\;l_k+l+j_{\gamma_t}+\lambda_t=\mbox{odd} \\
		0\;\mbox{if}\;l_k+l+j_{\gamma_t}+\lambda_t=\mbox{even} \\ 
	\end{array}
\right. ,
\end{eqnarray}
\begin{eqnarray}
\label{28}
\Delta^{j_{en},j_{en'}}(i,j,k)=\frac{(4\pi)^{3/2}\left[j_{ef},j_{en'},j_{en},j_{ei}\right]^{1/2}}{\left[j_{\gamma_i},j_{\gamma_j},j_{\gamma_k}\right]^{1/2}}
\begin{pmatrix}
j_{ef} & j_{\gamma_i} & j_{en'}\\
1/2 & 0 & -1/2
\end{pmatrix}
\begin{pmatrix}
j_{en'} & j_{\gamma_j} & j_{en}\\
1/2 & 0 & -1/2
\end{pmatrix}
\begin{pmatrix}
j_{en} & j_{\gamma_k} & j_{ei}\\
1/2 & 0 & -1/2
\end{pmatrix}
\Theta(i,j,k)
\end{eqnarray}
and 
\begin{eqnarray}
\label{29}
\Theta(i,j,k)=\left[j_{en'},j_{en}\right]^{1/2}\sum_{m_{en'},m_{en}}(-1)^{m_{ei}+m_{ef}+1}
\begin{pmatrix}
j_{ef} & j_{\gamma_i} & j_{en'}\\
-m_{ef} & m_{\gamma_i}  & m_{en'}
\end{pmatrix}
\begin{pmatrix}
j_{en'} & j_{\gamma_j}  & j_{en}\\
-m_{en'} & m_{\gamma_j}  & m_{en}
\end{pmatrix}
\begin{pmatrix}
j_{en} & j_{\gamma_k}  & j_{ei}\\
-m_{en} & m_{\gamma_k}  & m_{ei}
\end{pmatrix}\;.
\end{eqnarray}
The indices $i,j,k$ denote the serial number of the photon which can take the values $1,2,3$, the notation $[j,k, \dots ]$ means $(2j+ l)(2k+ 1)\dots$.

Finally expression for the decay rate can be written in the form
\begin{eqnarray}
\label{30}
\frac{dW_{i\rightarrow f}(\omega_1,\omega_2)}{d\omega_1d\omega_2} =\frac{\omega_3\;\omega_2\;\omega_1}{(2\pi)^5}\sum_{\lambda_1,\lambda_2,\lambda_3}\sum_{j_{\gamma_1}, j_{\gamma_2},j_{\gamma_3}}\sum_{m_{\gamma_1},m_{\gamma_2},m_{\gamma_3}}\left|\sum_{j_{en},j_{en'}}S^{j_{en'}j_{en}}(1,2,3)+(5\;\mbox{permutations})\right|^2\;.
\end{eqnarray}
Permutations in Eq. (\ref{30}) are understood as permutations of the indices $1,2,3$. 

The numerical results for the transition rates $2p_{1/2}\rightarrow 1s_{1/2}+3\gamma(E1)$ and $2s_{1/2}\rightarrow 1s_{1/2}+3\gamma(E2)$ are presented for the H-like ions with $ 1\leqslant Z \leqslant 95 $ in Tables I, II respectively. 
Summation over the full set of one-electron states was performed within the B-spline approach \cite{bspline}. The calculations were carried out in two relativistic "forms", corresponding to the nonrelativistic "length" and "velocity" forms \cite{labzsol}; the results coincide with 3-6 digits. All values was checked for the stability and convergence for the different length of the spline basis set. In order to integrate over photon frequency, the $ 24 $ points of Gauss-Legendre quadrature method was employed. For the summation over Dirac spectrum set from 40 B-spline basis states of order 9 were used. The frequency distributions of some transition probabilities are presented in Figs. 1-3.

\section{Three-photon transitions between fine structure components}
In this section we derive the expression for the 3-photon decay rate in H-like ion for a nuclei with nonzero spin $I$ and consider transitions between different hyperfine sublevels of different fine structure levels. The matrix element between states with total angular momentum $F_{\alpha}$ and $F_{\beta}$ (where $\left|j_e - I \right| \leqslant F \leqslant j_e + I$) can be reduces to the form \cite{varsh}:
\begin{eqnarray}
\label{37}
\left\langle n_{\alpha}F_{\alpha}M_{\alpha}\left|Q^{(\lambda)}_{j_{\gamma}m_{\gamma}\omega}\right|n_{\beta}F_{\beta}M_{\beta}\right\rangle = 
\\
\nonumber
(-1)^{F_{\alpha}-M_{\alpha}+j_{e{\alpha}+I}}\left[F_{\alpha}F_{\beta}\right]^{1/2}
\begin{pmatrix}
F_{\alpha} & j_{\gamma} & F_{\beta}\\
-M_{\alpha} & m_{\gamma} & M_{\beta}
\end{pmatrix}
\begin{Bmatrix}
F_{\alpha} & j_{\gamma} & F_{\beta}\\
j_{e\beta} & I  & j_{e\alpha}
\end{Bmatrix}
\left\langle n_{\alpha}j_{e\alpha}\left|\left|Q^{(\lambda)}_{j_{\gamma}\omega}\right|\right|n_{\beta}j_{e\beta}\right\rangle ,
\end{eqnarray}
\begin{eqnarray}
\label{38}
\left\langle n_{\alpha}j_{e\alpha}\left|\left|Q^{(\lambda)}_{j_{\gamma}\omega}\right|\right|n_{\beta}j_{e\beta}\right\rangle
=
%\\
%\nonumber
(-i)^{j_{\gamma}+\lambda-1}(-1)^{j_{e\alpha}-1/2}\left(\dfrac{4\pi}{2j_{\gamma}+1}\right)^{1/2}\left[j_{e\alpha},j_{e\beta}\right]^{1/2}
\begin{pmatrix}
j_{e\alpha} & j_{\gamma} & j_{e\beta}\\
1/2 & 0 & -1/2
\end{pmatrix} 
\overline{M}^{(\lambda,j_{\gamma})}_{\alpha\beta}(\omega).
\end{eqnarray}
Then the decay rate is
\begin{eqnarray}
\label{39}
\frac{dW_{if}}{d\omega_2d\omega_1}=\frac{\omega_3\omega_2\omega_1}{(2\pi)^5}\sum_{\lambda_1,\lambda_2,\lambda_3}\sum_{j_{\gamma_1}, j_{\gamma_2},j_{\gamma_3}}\sum_{m_{\gamma_1},m_{\gamma_2},m_{\gamma_3}}\left|\sum_{F_{e_{n'}},F_{e_n}}\sum_{j_{en'},j_{en}}S^{F_{n'}F_{n}j_{en'}j_{en}}(1,2,3)+(5\;\mbox{permutations})\right|^2\;,
\end{eqnarray}
where
\begin{eqnarray}
\label{44}
S^{F_{n'}F_{n}j_{en'}j_{en}}(i,j,k)=\sum_{l_{en'},l_{en}}\sum_{n,n'}\frac{\overline{M}^{(\lambda_i,j_{\gamma_i})}_{f,n'}(\omega_i)\overline{M}^{(\lambda_j,j_{\gamma_j})}_{n',n}(\omega_j)\overline{M}^{(\lambda_k,j_{\gamma_k})}_{n,i}(\omega_k)}{\left(E_{n'j_{en'}l_{en'}}-E_{n_fj_{ef}l_{ef}}-\omega_k-\omega_j\right)\left(E_{nj_{en}l_{en}}-E_{n_fj_{ef}l_{ef}}-\omega_k\right)}\times
\\
\nonumber
\Delta^{F_{n'},F_{n},j_{en'},j_{en}}(i,j,k)\;\pi^{l_{en'}}_f(i)\;\pi^{l_{en}}_{n'}(j)\;\pi^{l_{ei}}_{n}(k),
\end{eqnarray}
\begin{eqnarray}
\label{46}
\Delta^{F_{n'},F_{n},j_{en'},j_{en}}(i,j,k)=\frac{(4\pi)^{3/2}\left[j_{ef},j_{en'},j_{en},j_{ei}\right]^{1/2}}{\left[j_{\gamma_i},j_{\gamma_j},j_{\gamma_k}\right]^{1/2}}
\begin{pmatrix}
j_{ef} & j_{\gamma_i} & j_{en'}\\
1/2 & 0 & -1/2
\end{pmatrix}
\begin{pmatrix}
j_{en'} & j_{\gamma_j} & j_{en}\\
1/2 & 0 & -1/2
\end{pmatrix}
\\
\nonumber
\times
\begin{pmatrix}
j_{en} & j_{\gamma_k} & j_{ei}\\
1/2 & 0 & -1/2
\end{pmatrix}
\left[j_{en'},j_{en}\right]^{1/2}\left[F_{n'},F_{n}\right]^{1/2}
\begin{Bmatrix}
F_{f} & j_{\gamma_i} & F_{n'}\\
j_{en'} & I  & j_{ef}
\end{Bmatrix}
\begin{Bmatrix}
F_{n'} & j_{\gamma_j} & F_{n}\\
j_{en} & I  & j_{en'}
\end{Bmatrix}
\begin{Bmatrix}
F_{n} & j_{\gamma_k} & F_{i}\\
j_{ei} & I  & j_{en}
\end{Bmatrix}\Theta(i,j,k)
\end{eqnarray}
and
\begin{eqnarray}
\label{46a}
\Theta(i,j,k)=\left[F_{n'},F_{n}\right]^{1/2}\sum_{M_{n'},M_{n}}(-1)^{M_{f}+M_{n'}+M_{n}}
\begin{pmatrix}
F_{f} & j_{\gamma_i} & F_{n'}\\
-M_{f} & m_{\gamma_i} & M_{n'}
\end{pmatrix}
\begin{pmatrix}
F_{n'} & j_{\gamma_j} & F_{n}\\
-M_{n'} & m_{\gamma_j} & M_{n}
\end{pmatrix}\\
\nonumber\times
\begin{pmatrix}
F_{n} & j_{\gamma_k} & F_{i}\\
-M_{n} & m_{\gamma_k} & M_{i}
\end{pmatrix}\;.
\end{eqnarray}
In Eq. (\ref{39}) we have neglected the energy shift due to the hyperfine splitting, since it is small with respect to the total energy difference. In Table III the numerical results for transition probabilities of $2p_{3/2}(F=0)\rightarrow 2s_{1/2}(F=2)+3\gamma (E1)$  decay in H-like ions for the arbitrary nuclear charge $Z$ and nuclear spin $I=3/2$ are listed. In the two last columns of Table III the hyperfine splitting coefficients $ A(\alpha Z) $ for the levels $ 2p_{3/2} $ and $ 2s_{1/2} $ for different $ Z $ values are given. These splittings are evaluated in a fully relativistic theory according to \cite{shabaev}
\begin{eqnarray}
\label{46b}
\Delta E_{\mu}= A(\alpha Z)\frac{F(F+1)-I(I+1)-j_e(j_e+1)}{2Ij_e(j_e+1)(2l_e+1)}\;,
\end{eqnarray}   
\begin{eqnarray}
\label{46c}
A(\alpha Z)=\alpha(\alpha Z)^3\frac{\mu}{\mu_{N}}\frac{m_e}{m_p}\frac{(2l_e+1)\kappa\left[2\kappa(\gamma +n_r)-N\right]}{N^4\gamma(4\gamma^2-1)}\;.
\end{eqnarray}
Here $ I $ is the nuclear spin, $ \kappa=(-1)^{j_e+l_e+1/2}(j_e+1/2) $ is a relativistic electron angular quantum number, $ n_r $ is the radial quantum number ($ n_r=n-|\kappa| $), $ \gamma=\sqrt{\kappa^2-(\alpha Z)^2} $, $ N=\sqrt{n_r^2+2n_r\gamma+\kappa^2} $ and $ \mu $ is the magnetic moment of nucleus expressed in units of nuclear magneton $ \mu_N=|e|\hslash/(2m_pc) $, $ m_e $ and $ m_p $ is the electron and proton mass, respectively. The values of $ I $, $ \mu $ are also given in Table III.

These transitions can be interesting in the view of the possible tests of Bose-Einstein statistics for the multiphoton systems \cite{zlsg}.

\section{Application for the test of Bose-Einstein statistics}
In \cite{zlsg} the SSSRs for the multiphoton atomic transitions with equivalent photons which present an extension of the LYT \cite{landau}, \cite{yang} were formulated. These rules consist of: 1) SSSR-1: Two equivalent photons involved in any atomic transition can have only even values of the total angular momentum $J$, 2) SSSR-2: Three equivalent dipole photons involved in any atomic transition can have only odd values of the total angular momentum $J=1, \,3$, 3) SSSR-3: Four equivalent dipole photons involved in any atomic transition can have only even values of the total momentum values $J=0,\,2,\, 4$. It was established in \cite{zlsg} that SSSR-2, SSSR-3 do not hold, in general, for the photon multipolarity $j>1$. 

In \cite{zlsg} the experiments on the test of SSSRs with the absorption processes in He-like ion of Uranium were suggested where the lasers can be used as a  source for the the equivalent photons. An advantage of the use of the laser source is that all the photons will have the same frequency. If we divide this frequency by an integer number $N_{\gamma}$ and adjust the laser frequency $\omega_l$ to the value of transitions frequency $ \omega_a $, $\omega_l=\omega_a/N_{\gamma}$, the number of photons $N_{\gamma}$ in the absorption process will be fixed. 

Using transitions with the initial total electron momentum $ J_i=2 $ and the final momentum $ J_f=0 $ we will fix the total momentum of the photon system $ J $ equal to $ J_i $. Then, choosing $ N_{\gamma}=3 $ and $ J_i=2 $ we will test SSSR-2. According to SSSR-2 the value $ J=2 $ for 3 equal photons is forbidden, so that the absorption of the laser light at the corresponding frequency $ \omega_l=\omega_a/3 $ should be absent. In the same way SSSR-3 can be tested. The numerical examples with the highly charged He-like ions were given in \cite{zlsg}. The photon frequencies in this case are in the X-ray region.

The experiments of this type can be extended to the H-like ions with the half-integer nuclear spin $ I $. The advantage of such experiments consists in the possibility to use the optical range lasers. In the recent experiments with heavy atoms and ions \cite{mokler} it is possible to measure the frequency distribution for the transitions rates. In this case the value of the total angular momentum $F$ for $N_\gamma$-photon system can be fixed by choosing the appropriate values $F_{i}$ and $F_{f}$ for the initial (lower) and final (upper) levels in the transition process. 

In laser beam all the possible multipolarities of photon are presented. Thus it should produce all the transitions with the same total parity: E1E1E1, E1M1E2, E1E1M2 etc. However the processes with the photons of higher multipolarities are usually strongly suppressed in atoms. Due to this suppression the E1E1E1 transition will be dominant. Measuring the absorption rate at the $\omega_l=\omega_a/N_{\gamma}$ frequency one can establish the validity or non-validity of the particular SSSRs: the atomic vapour should be transparent for the laser light at the frequency $\omega_l=\omega_a/N_{\gamma}$. Note also that unlike the spontaneous emission which is very weak for multiphoton transitions, the multiphoton absorption depends on the laser intensity and can be well observed in the experiments.

In our examples we considered transitions between fine structure components with fixation of a certain fine structure subcomponents. Varying the nuclear charge $Z$ we can find the situation when each photon will be in optical range ($0.857$ eV - $3.27$ eV). For example, to test the SSSRs the H-like ions with $ Z=16,\;17,\;19 $ can be examined. In this case transitions between fine structure components $  2s_{1/2}(F=2) \rightarrow 2p_{3/2}(F=0)+3\gamma(E1) $ can be chosen. The energy intervals between $ 2p_{3/2}(F=0) $ and $ 2s_{1/2}(F=2) $ states for $ Z=16,\;17,\;19 $ are listed in the 8th column of Table III and corresponding $ \omega_l$ for $ N_{\gamma}=3 $ is in the optical region. The nuclei of these ions are stable \cite{magneton}. It is important that for $ Z=16,\;17,\;19 $ the hyperfine splitting both for $ 2p_{3/2} $ and $ 2s_{1/2} $ are resolvable (see Table III) i.e. the transition $ 2s_{1/2}(F=2)\rightarrow 2p_{3/2}(F=0) $ can be well separated out.  The $ 3E1 $ transition rate value is $ 10^{13} $ times smaller then the one-photon M2 transition rate. However tuning the laser frequency to the one third of transition frequencies excludes one-photon absorption. Thus the frequency distribution depicted in Fig. 1 (and its two-dimensional sectional cut Fig. 2) should be observable in experiments of such type. The same picture arises for $ 1s_{1/2}(F=0)\rightarrow 2p_{3/2}(F=0)+3\gamma(E1) $ transition in neutral hydrogen atom ($ Z=1 $) with $ I=1/2 $ and $ \mu=2.793 $ (see Fig. 3). In this case $ \omega_l=3.40148 $ eV. For this transition the virtual states $ ns_{1/2}(F=1) $ and $ np_{1/2}(F=1) $ in the sum in Eq. (\ref{39}) for the $ n=2 $ lie between the initial $ 2p_{3/2}(F=2) $ and final $ 1s_{1/2}(F=0) $ states which leads to the resonance (the situation when the energy denominator turns to zero). The presence of the cascade-producing states in the sum over the intermediate states in the transition amplitude leads to the arrival of the high, but narrow "ridge" in the frequency distribution $\frac{dW(\omega_1,\omega_2)}{d\omega_1d\omega_2}$ \cite{zlsg}. The "ridge" does not influence the SSSR-2: it does not correspond to the case of three equivalent photons. This is a general situation for all the possible cascade transitions. While in \cite{demille, english} it was demonstrated that two photons behave like two bosons the experiments suggested above would demonstrate that three photons also obey the Bose-Einstein statistics. 

\begin{center}
Acknowledgments
\end{center}
The work was supported by RFBR (grants No. 14-02-00188). T. Z., D. S. and L. L. acknowledge the support by St.-Petersburg State University with a research grant 11.38.227.2014. The work of T. Z. was supported also by the nonprofit foundation
"Dynasty" (Moscow).

\begin{table}[hbtp]
\caption{Transition probabilities for $2p_{1/2}\rightarrow 1s_{1/2}+3\gamma(E1)$ and $2p_{1/2}\rightarrow 1s_{1/2}+\gamma(E1)$ decay rates in $s^{-1}$ for different $Z$. The number in parentheses indicates the power of ten. Transition energies in eV are listed in the last column.}
\begin{tabular}{ c  c  c  c  c}
\\
\hline
 $ Z $  & $ W^{3\gamma}_{vel.} $ & $ W^{3\gamma}_{len.} $ & $ W^{1\gamma} $ & $ \Delta E $
\qquad\\
\hline 
 $ 1 $ & $ 1.168620(-8)  $ & $ 1.168632(-8) $ & $ 6.268(8)$ & $ 10.204393  $
\qquad\\
 $ 5 $ & $ 4.561173(-3)  $ & $ 4.561219(-3) $ & $ 6.918(11) $ & $  2.551846(2) $
\qquad\\
 $ 10 $ & $ 1.164659  $ & $ 1.164670 $ & $ 6.272(12) $ & $ 1.021675(3)  $
\qquad\\
 $ 20 $ & $ 2.950782(2) $ & $ 2.950754(2) $ & $ 1.005(14)  $ & $ 4.101827(3) $
\qquad\\
 $ 30 $ & $ 7.430658(3)  $ & $ 7.430721(3) $ & $ 5.106(14) $ & $ 9.287031(3) $
\qquad\\
 $ 40 $ & $ 7.237156(4)  $ & $ 7.237208(4) $ & $ 1.621(15) $ & $  1.665912(4) $
\qquad\\
 $ 50 $ & $  4.170711(5) $ & $ 4.170735(5) $ & $ 3.980(15) $ & $ 2.634226(4)  $
\qquad\\
 $ 60 $ & $ 1.717285(6)  $ & $ 1.717293(6)  $ & $ 8.310(15) $  & $ 3.851493(4) $
\qquad\\
 $ 70 $ & $ 5.579235(6) $ & $  5.579255(6)  $ & $ 1.552(16) $  & $ 5.342837(4) $
\qquad\\
 $ 80 $ & $ 1.514879(7) $ & $ 1.514883(7)  $ & $ 2.671(16) $ & $ 7.144008(4) $
\qquad\\
 $ 90 $ & $ 3.554091(7) $ & $ 3.554098(7) $ &  $ 4.316(16) $ & $ 9.305821(4)  $
\qquad\\
 $ 95 $ & $ 5.1821200(7)$ & $ 51821200(7) $ & $ 5.380(16) $ & $  1.054361(5) $
\qquad\\
\hline
\end{tabular}
\end{table}

\begin{table}[hbtp]
\caption{Transition probabilities for $2s_{1/2}\rightarrow 1s_{1/2}+3\gamma(E2)$ and $2s_{1/2}\rightarrow 1s_{1/2}+2\gamma(E1)$ in $s^{-1}$ for different $Z$. The number in parentheses indicates the power of ten. Transition energies in eV are listed in the last column.}
\begin{tabular}{ c  c  c  c  c}
\\
\hline
 $ Z $  & $ W^{3\gamma}_{vel.} $ & $ W^{3\gamma}_{len.} $ & $ W^{2\gamma} $ & $ \Delta E $
\qquad\\
\hline 
 $ 1 $ & $ 1.284270(-27) $ & $ 1.284254(-27) $ & $ 8.229063 $ & $ 10.204393  $
\qquad\\
 $ 5 $ & $ 7.833773(-18) $ & $ 7.833681(-18) $ & $ 1.284705(5) $ & $ 2.551846(3) $
\qquad\\
 $ 10 $ & $ 1.281043(-13) $ & $ 1.284925(-13) $ & $ 8.200646(6) $ & $ 1.021674(3) $
\qquad\\
 $ 20 $ & $ 2.082775(-9) $ & $ 2.087035(-9) $ & $ 5.195127(8)  $ & $ 4.101829(3) $
\qquad\\
 $ 30 $ & $ 6.001701(-7) $ & $ 6.008097(-7) $ & $ 5.821090(9)  $ & $ 9.287046(3) $
\qquad\\
 $ 40 $ & $ 3.306380(-5) $ & $ 3.307976(-5) $ & $ 3.198616(10) $ & $  1.665919(4) $
\qquad\\
 $ 50 $ & $ 7.335633(-4) $ & $ 7.337452(-4) $ & $ 1.186615(11) $ & $ 2.634252(4) $
\qquad\\
 $ 60 $ & $ 9.129899(-3) $ & $ 9.131310(-3) $ & $ 3.426453(11) $ & $ 3.851581(4) $
\qquad\\
 $ 70 $ & $ 7.601921(-2) $ & $ 7.602718(-2) $ & $ 8.305995(11) $ & $ 5.343122(4)  $
\qquad\\
 $ 80 $ & $ 0.470047 $ & $ 0.470080 $ & $ 1.767273(12) $ & $ 7.144853(4)  $
\qquad\\
 $ 90 $ & $ 2.304994 $ & $ 2.305110 $ & $ 3.393473(12) $ & $ 9.308396(4)  $
\qquad\\
 $ 95 $ & $ 4.744875 $ & $ 4.745083 $ & $ 4.551134(12) $ & $ 1.054819(5) $
\qquad\\
\hline
\end{tabular}
\end{table}

\begin{table}[hbtp]
\caption{Transition probabilities for $2p_{3/2}(F=0)\rightarrow 2s_{1/2}(F=2)+3\gamma(E1)$ and $ 2p_{3/2}(F=0)\rightarrow 2s_{1/2}(F=2)+\gamma(M2) $  in $s^{-1}$ for different H-like ions. Nuclear spin $ I=3/2 $. The number in parentheses indicates the power of ten. Transition energies and hyperfine splitting constants $ A $ (Eq. (\ref{46c})) in eV are listed in three last columns.}
\begin{tabular}{c  c  c  c  c  c  c  c  c  c}
\\
\hline
 $ \mbox{Ion} $ & $ \mbox{Abundance}\;\% $ & $ \mu $ & $ I $ & $ W^{3\gamma}_{vel.} $ & $ W^{3\gamma}_{len.} $ & $ W^{1\gamma} $ & $ \Delta E $ & $ A_{2p_{3/2}} $ & $ A_{2s_{1/2}} $
\qquad\\
\hline
 $ \;^{33}\mbox{S}^{15+} $ & $ 0.75 $ & $ 0.64382 $ & $ \frac{3}{2} $ & $ 8.944609\times 10^{-20} $ & $ 8.944608\times 10^{-20} $ & $ 6.406678\times 10^{-6} $ & $ 2.992524 $ & $ 2.61179\times 10^{-4} $ & $ 2.67850\times 10^{-4} $
\qquad\\
 $ \;^{35}\mbox{Cl}^{16+} $ & $ 93.3 $ & $ 0.82187 $ & $ \frac{3}{2} $ & $ 3.404860\times 10^{-19} $ & $ 3.404860\times 10^{-19} $ & $ 1.917035\times 10^{-5} $ & $ 3.817932 $ & $ 4.00118\times 10^{-4} $ & $ 4.11687\times 10^{-4} $
\qquad\\
 $ \;^{39}\mbox{K}^{18+} $ & $ 75.8 $ & $ 0.39147 $ & $ \frac{3}{2} $ & $ 3.963697\times 10^{-18} $ & $ 3.963697\times 10^{-18} $ & $ 1.434485\times 10^{-4} $ & $ 5.971640 $ & $ 5.59230\times 10^{-4} $ & $ 5.79554\times 10^{-4} $
\qquad\\
\hline
\end{tabular}
\end{table}

\begin{figure}[hbtp]
\caption{(Color Online) 3-dimensional plot for frequencies distribution of the transition rate $ 2p_{3/2}(F=0)\rightarrow 2s_{1/2}(F=2)+3\gamma(E1) $ in H-like ion with $ Z=19 $. Nuclear spin $ I=3/2 $. On the vertical axis the transition rate $ \frac{dW}{d\omega_1d\omega_2}$ in $ s^{-1} $ is plotted; on the horizontal axes the photon frequencies are plotted in units $ \omega_1/\Delta $, $ \omega_2/\Delta  $ where  $ \Delta $ denotes the energy difference $ \Delta=E(2p_{3/2})-E(2s_{1/2}) $. The lowest (zero) point is the point with coordinates $ \omega_1/\Delta=\omega_2/\Delta=1/3 $ at the bottom of the "pit" in the frequency distribution for the transition rate which arises due to SSSR-2. }
\centering
\includegraphics[scale=0.55]{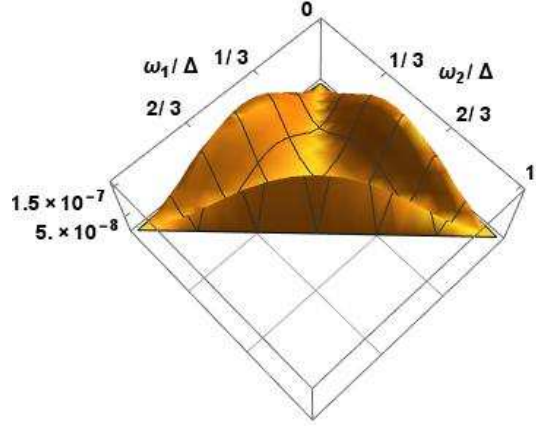}
\end{figure}

\begin{figure}[hbtp]
\caption{(Color Online) Two-dimensional sectional cut of Fig. 1 at the frequency point $ \omega_2/\Delta  =1/3$.  All details are the same as in Fig. 1. The lowest (zero) point is the point with coordinates $ \omega_1/\Delta=1/3 $ arises due to SSSR-2.}
\centering
\includegraphics[scale=0.55]{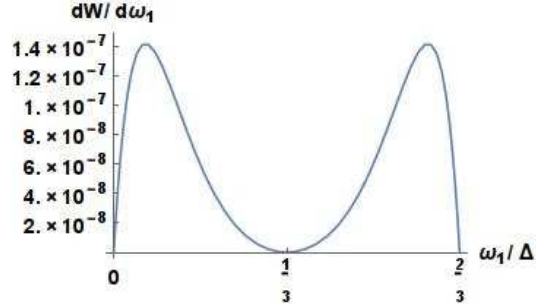}
\end{figure}

\begin{figure}[hbtp]
\caption{(Color Online) 3-dimensional plot for frequencies distribution of the transition rate $ 2p_{3/2}(F=2)\rightarrow 1s_{1/2}(F=0)+3\gamma(E1) $ in the neutral hydrogen atom ($ Z=1 $). The energy difference between $ 2p_{3/2}(F=2) $ and $ 1s_{1/2}(F=0) $ states is 10.204435 eV and $ \omega_l=3.40148 $ eV. All details are the same as in Fig. 1. Here $ \Delta=E(2p_{3/2})-E(1s_{1/2}) $. The lowest (zero) point is the point with coordinates $ \omega_1/\Delta=\omega_2/\Delta=1/3 $ at the bottom of the "pit" in the frequency distribution for the transition rate which arises due to SSSR-2. }
\centering
\includegraphics[scale=0.55]{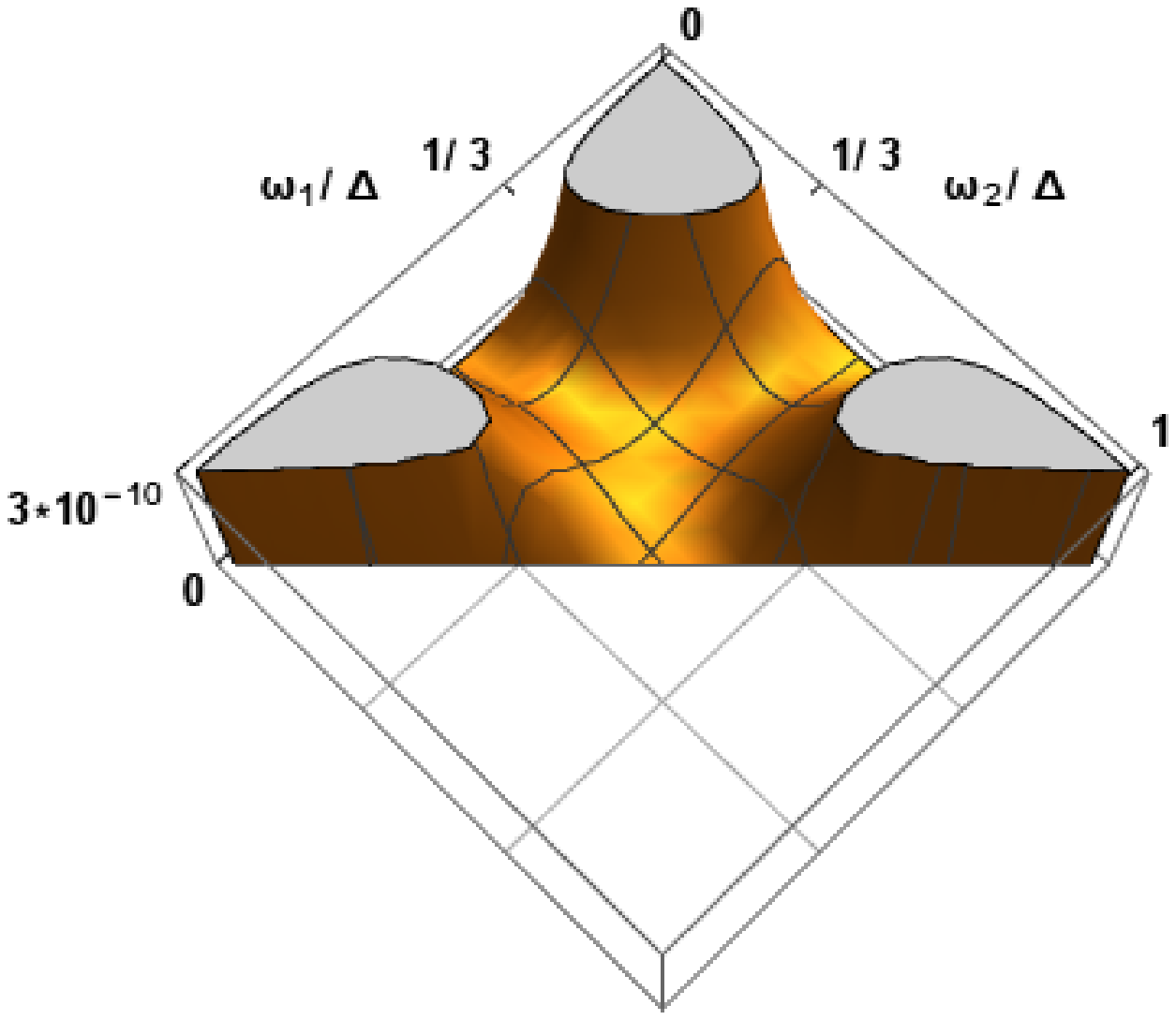}
\end{figure}


\begin{thebibliography}{99}
\bibitem{kramers} H. H. Kramers and W. Heisenberg, Z. Phys. {\bf 31}, 681 (1925).

\bibitem{waller} I. Waller, Z. Phys. {\bf 58}, 75 (1928).

\bibitem{goeppert} M. Goeppert-Mayer, Ann. Phys. (Leipzig) {\bf 9}, 273 (1931).

\bibitem{seager} S. Seager, D. D. Sasselov, and D. Scott, Astrophys. J. Lett. {\bf 523}, L1 (1999).

\bibitem{chluba} J. Chluba and R. A. Sunyaev, Astron. Astrophys. {\bf 446}, 39 (2006).

\bibitem{demille} D. DeMille, D. Budker, N. Derr and E. Deveney, Phys. Rev. Lett. {\bf 83}, 3978 (1999).

\bibitem{english}  D. English, V. V. Yashchuk and D. Budker, Phys. Rev. Lett. {\bf 104}, 253604 (2010).

\bibitem{dunford}  R. W. Dunford, Phys. Rev. A {\bf 69}, 062502 (2004).

\bibitem{kozlov} M. G. Kozlov, D. English and D. Budker, Phys. Rev. A {\bf 80}, 042504 (2009).

\bibitem{angom} D. Angom, K. Bhattacharya, S. D. Rindani, Int. J. Mod. Phys. A {\bf 22}, 707 (2007).

\bibitem{bspline} W. R. Johnson, S. A. Blundell, and J. Sapirstein, Phys. Rev. A {\bf 37}, 307 (1988).

\bibitem{zlsg} T. Zalialiutdinov, D. Solovyev, L. Labzowsky and G. Plunien
Phys. Rev. A {\bf 91}, 033417 (2015).

\bibitem{landau} L. D. Landau, Dokl. Akad. Nauk SSSR {\bf 60}, 207 (1948).

\bibitem{yang} C. N. Yang, Phys. Rev. {\bf 77}, 242 (1950).

\bibitem{berlif} V. B. Berestetskii, E. M. Lifshitz and L. P. Pitaevskii, {\it Quantum Electrodynamics}, Oxford, Pergamon, 1982.

\bibitem{akhiezer} A. I. Akhiezer and V. B. Berestetskii, {\it Quantum Electrodynamics}, New York, Wiley 1965.

\bibitem{ALPS} O. Yu. Andreev, L. N. Labzowsky, G. Plunien and D. A. Solovyev, Phys. Rep. {\bf 455}, 135 (2008).

\bibitem{grant} I. P. Grant, J. Phys. B: Atom. Mol. Phys. {\bf 7}, 1458 (1974). 

\bibitem{goldman} S. P. Goldman and G. W. F. Drake, Phys. Rev. A {\bf 24}, 183 (1981).

\bibitem{labzsol} L. Labzowsky, D. Solovyev, G. Plunien and G. Soff, Eur. Phys. J. D {\bf 37}, 335 (2006).

\bibitem{varsh} D. A. Varshalovich, A. N. Moskalev and V. K. Khersonskii, {\it Quantum Theory of Angular Momentum}, World Scientific, Singapore (1988).

\bibitem{shabaev} V. M. Shabaev, J. Phys. B {\bf 27}, 5825 (1994).

\bibitem{mokler} P. H. Mokler and R. W. Dunford, 2004, Phys. Scr. {\bf 69} C1.

\bibitem{magneton} N. J. Stone, Atomic Data and Nuclear Data Tables, {\bf 90},  pp. 75-176 (2005).

\end{thebibliography}
\end{document}